\documentclass[fontsize=12pt, paper=a4, headinclude, twoside=true, parskip=half+, pagesize=auto, numbers=noenddot, open=right, toc=listof, toc=bibliography]{article}
\usepackage{amsmath}
\usepackage{amsthm}
\usepackage{amsfonts}

\usepackage{wrapfig}
\usepackage{graphicx}
\usepackage{caption}
\usepackage{subcaption}

\usepackage{cleveref}
\Crefname{equation}{Eq.}{Eqs.}
\Crefname{figure}{Fig.}{Figs.}
\Crefname{tabular}{Tab.}{Tabs.}

\crefname{equation}{eq.}{eqs.}
\crefname{figure}{fig.}{figs.}
\crefname{tabular}{tab.}{tabs.}
\usepackage{url}

\usepackage[utf8]{inputenc}
\usepackage{xcolor}
\begin{document}
	\title{Hierarchical clusters in neuronal populations with plasticity}
	\author{Vera R\"ohr$^{1}$, Rico Berner$^{2,3}$, Ewandson L. 
Lameu$^{4,5}$, \\ Oleksandr V. Popovych$^{6,7}$, Serhiy Yanchuk$^3$
}
	\maketitle
{\centering\small
$^1$ Neurotechnology Group, Technische Universit{\"a}t Berlin;
$^2$ Institut f{\"u}r Theoretische Physik, Technische Universit{\"a}t Berlin;
$^3$ Institut f{\"u}r Mathematik, Technische Universit{\"a}t Berlin;
$^4$ National Institute for Space Research (INPE), Brazil;
$^5$ Institut f{\"u}r Physik, Humboldt-Universit{\"a}t zu Berlin
$^6$ Institute of Neuroscience and Medicine - Brain $\&$ Behaviour
(INM-7), Research Centre J\"{u}lich, Germany; 
$^{7}$Institute for Systems Neuroscience, Medical Faculty, Heinrich-Heine University
D\"{u}sseldorf,  Germany
}

\section*{Abstract}
We report the phenomenon of frequency clustering in
a network of Hodgkin-Huxley neurons with spike
timing-dependent plasticity. The clustering leads to a splitting of 
a neural population into a few groups synchronized at different frequencies.
In this regime, the amplitude of the mean field undergoes low-frequency
modulations, which may contribute to the mechanism of the emergence of slow 
oscillations of neural activity observed  
in spectral power of local field potentials or electroencephalographic
signals at high frequencies. 
In addition to numerical simulations of such multi-clusters, we investigate the 
mechanisms of the observed phenomena using the simplest case of two clusters. 
In particular, we propose a phenomenological model which describes the dynamics 
of two clusters taking into account the adaptation of coupling weights.  
We also determine the set of plasticity functions (update rules), 
which lead to multi-clustering.

\section*{Author summary}

Synaptic plasticity is one of the key mechanisms that allow for neural networks 
to adapt their structure. Depending on the mutual neural activity, the efficacy 
of synapses may change, which results in short- or long-term potentiation or depression 
of a synapse. In this paper we investigate the structural changes that are caused 
by the spike timing-dependent plasticity (STDP), where the synaptic weights 
are adapted depending on the difference of spiking times between the pre- and 
postsynaptic neuron. The adaptation is considered to be symmetric as experimentally 
found for hippocampal synapses \cite{Wittenberg2006} and can also be derived from asymmetric STDP under certain conditions \cite{Kepecs2002}.
In an adaptive network of Hodgkin-Huxley neurons 
we observe the emergence of clusters of neurons that are synchronized 
at different frequencies in different clusters. Our study shows that such 
a self-organized cluster formation is robust against changes of plasticity function. 
While the spiking frequency of each synchronized cluster appear to be on the 
timescale of individual neurons, the amplitude of the mean field of such cluster 
system can evolve at a few orders of magnitude slower timescales. 
The reported slow modulations of the mean neural activity in 
a neural population with plasticity can explain the emergence of slow cortical 
oscillations detected in the empirical data of spectral power of local field
potentials (LFP) and electroencephalographic (EEG) signals, which correlate 
with spontaneous fluctuations of the blood oxygen level-dependent (BOLD) signals 
measured by functional magnetic resonance imaging (fMRI) \cite{Mantini2007,Magri2012}.

\section*{Introduction}
Clustering of the dynamics and coupling is
observed at several scales of the brain structure
and function. For example, in the data measured by the functional
magnetic resonance imaging (fMRI), the brain networks form functional
clusters that can be seen in the matrices of the functional and effective
connectivities for task-based and task-free (resting state) paradigms
\cite{Leonardi2013,Cole2014,Marrelec2016,Bolt2017,Zhou2018, Fornito2012}. Clustering has
also been observed for dynamic functional connectivity, where the
time courses of the connectivity exhibit a few discrete states with
well pronounced clusters \cite{Allen2014}. 
Disruption of such clustered states of connectivity may be associated
with some brain disorders \cite{Damaraju2014,Uhlhaas2006}. 
It is therefore important to investigate the emergence of clustering in neural populations
that we address in this study.

Neural networks are able to adapt their structure depending on the activity of 
the nodes or external stimuli \cite{Gerstner2014}.
One of the possible mechanisms of such an adaptation, which may lead to persistent 
changes in neural connections and relate to learning and memory, 
is synaptic plasticity \cite{Hebb1949}. The efficacy of synapses to transmit 
the electrical potential between neurons may increase or decrease depending on 
the mutual neural activity, which results in short- or long-term potentiation 
or, respectively, depression of synapses \cite{Brown1988,Bliss1993}.
An example is spike timing-dependent plasticity (STDP) which describes 
the synaptic weight change as a function of the difference of spiking 
times between pre- and post-synaptic neurons 
\cite{Gerstner1996,Markram1997a,Bi1998,Abbott2000,Bi2001}. 

One of the famous plasticity rules, the Hebbian rule, assumes that the 
modifications of the synaptic weights are driven by correlations in the firing 
activity of pre- and post-synaptic neurons. More specifically, it assumes that 
those connections are potentiated, for which one neuron contributes to the 
firing of another 
\cite{Hebb1949}. Nevertheless, in many publications, the Hebbian rule is 
considered in a more narrow sense of a closeness between the spiking times: 
the smaller the distances between the spikes are the higher is the potentiation 
of  the corresponding synapse \cite{Hoppensteadt1996,Seliger2002}. In this 
work, we are dealing with spike-based learning rules rather than rate-based. 

Previous studies of neural networks with STDP showed that such networks can 
evolve and create various coupling structures. For instance, the weights can 
exhibit  stable localized spatial structures, that can be interpreted as 
receptive fields \cite{Clopath2010}. These structures can be either 
unidirectionally of bidirectionally coupled, depending on the plasticity rule 
or external input properties. The STDP mechanism plays an important role in 
temporal coding of information by spikes \cite{Gerstner1996,Clopath2010}. 
On the one hand, a synchronized firing in neural ensembles with STDP can be  
stabilized through potentiation of synaptic coupling by stimulation-induced 
transient synchronization of neurons \cite{Tass2012b,Tass2006a,Popovych2013,Lucken2016}. 
On the other hand, a desynchronized state can lead to a depression of synaptic weights 
\cite{Tass2012b,Tass2006a}. Thus, neural networks with plasticity are prone 
to a co-existence of different stable dynamical and structural states, 
which may be realized by choosing appropriate initial conditions 
or stimulation procedures. 

Human brain networks demonstrate different degrees of modularity, sometimes with hierarchical features \cite{ref1, Bassett9239, Bassett2, Betzel, Lohse}.
Recently, a hierarchical clustering was observed in phenomenological models of 
adaptive networks of phase oscillators 
\cite{Aoki2011,Kasatkin2017,Berner2018}. As a result of an 
adaptation, the network evolved into groups of strongly-connected clusters, 
while the coupling between the groups was depressed. The stability analysis of 
such clusters reveals \cite{Berner2018,Berner2019} that the preferred stable cluster 
configuration corresponds to significantly different sizes of the clusters. The 
dynamics within each cluster are frequency-synchronized, while the frequencies 
between clusters differ. Thus, self-organized emergence of clusters leads to 
the emergence of different collective frequencies in the system. The 
multi-stability of such clusters was also observed in ensembles of Morris-Lecar 
bursting neurons with STDP in \cite{Popovych2015}.

In this paper we report on the phenomenon of clustering 
with respect to connectivity and frequencies in a network 
of adaptively coupled Hodgkin-Huxley (HH) neurons. The spike timing-dependent 
adaptation is considered to be symmetric as experimentally found for 
hippocampal synapses \cite{Wittenberg2006} and can also be derived 
from asymmetric STDP for an "effective time window" \cite{Kepecs2002}. Then 
the observed clusters are bidirectionally coupled \cite{Popovych2015}. Splitting 
of a neural population to a few clusters synchronized at different frequencies  
could lead to a slow waxing and waning of the amplitude of the mean field, 
where the clusters transiently gather together and move apart as the time 
evolves \cite{Popovych2015}. The frequency of such a modulation of the mean 
neural activity could be much smaller than the firing rate of individual 
neurons and depends on the differences between the clusters' frequencies. 
The emergence of synchronized clusters could explain the origin of the 
low-frequency modulation of the spectral power of macroscopic brain signals 
like local field potentials (LFP) or electroencephalographic (EEG) signals 
in higher frequency bands, which also correlates with slow oscillations of 
the blood oxygen level-dependent (BOLD) signal measured by fMRI \cite{Mantini2007,Magri2012, Monto2008,Alvarado2014}. 
Several other modeling studies have
also reported on clustering in the neural populations with plasticity
\cite{Maistrenko2007,Cateau2008,Popovych2015}. These clusters have
been observed for different models that ranged from simple phase oscillators
to the models of spiking and bursting neurons and demonstrate stability
with respect to heterogeneity of the interacting neurons and random
perturbations \cite{Maistrenko2007,Cateau2008,Popovych2015}.
In this paper we provide a simple phenomenological model and 
explain a mechanism governed by synaptic plasticity of 
the stabilization of such clusters in 
a neural population. 

The structure of the paper is as follows. In 
the first section we present the model. The next section shows 
numerically observed multi-clusters.
The detailed mechanisms of the stability of frequency 
clusters is explained afterwards using the simplest case of 
two clusters. Then we propose a phenomenological model, 
which describes the dynamics of two clusters taking into account the adaptation 
of the weights. 
The model is shown to reflect not only qualitative, 
but also some basic quantitative properties of 
the two-cluster formation. We also determine the set 
of plasticity functions (update rules), which lead to 
the clustering.  

\section*{Materials and methods}
\subsection*{Model\label{sec-model}}
The network of $N$ HH neurons is described by the following system 
\cite{Hodgkin1952,Hansel1993,Popovych2013,Lucken2016}
\begin{equation} \label{firsteq}
\begin{split}
C\dot{V}_i & =I_{i}-g_{Na}m_i^3h_i(V_i-E_{Na})-g_kn_i^4(V_i-E_K)-g_L(V_i-E_L) 
\\
 & \hspace{6cm} -\frac{(V_i-E_{r})}{N}\sum_{j=1}^{N}\kappa_{ij}s_j ,\\ 
\dot{m}_i & =\alpha_m(V_i)(1-m_i)-\beta_m(V_i)m_i ,\\
\dot{h}_i & =\alpha_h(V_i)(1-h_i)-\beta_h(V_i)h_i ,\\
 \dot{n}_i & =\alpha_n(V_i)(1-n_i)-\beta_n(V_i)n_i ,\\
  \dot{s}_i  & = \frac{5(1-s_i)}{1+e^(\frac{-V_i+3}{8})}-s_i .
\end{split}
\end{equation}
Here $V_i$ is the potential of the $i$-th neuron with the corresponding 
equilibrium potentials $E_{Na}=50$mV, $E_{K}=-77$mV, and $E_{l}=-54.4$mV. 
$C=1\mu F/cm^2$. Our choice of $E_{r}=20$mV corresponds to the excitatory neurons. 
$m$, $h$, and $n$ 
are gating variables, and their dynamics depend on opening and closing rates 
\begin{align*}
\alpha_m(V)&=\frac{0.1V+4}{1-e^{(-0.1V-4)}},\\
\beta_m(V)&=4e^{(\frac{-V-65}{18})},\\
\alpha_h(V)&=0.07e^{(\frac{-V-65}{20})},\\
\beta_h(V)&=\frac{1}{1+e^{(-0.2V-3.5)}},\\
\alpha_n(V)&=\frac{0.01V+0.55}{1-e^{(-0.1V-5.5)}},\\
\beta_n(V)&=0.125e^{(\frac{-V-65}{80})}.
\end{align*}
The parameters are  $g_{Na}=120${mS}/{cm$^2$}, $g_{K}=36${mS}/{cm$^2$}, and 
$g_{l}= 0.3${mS}/{cm$^2$}.
The constant current $I_{i}$ is set to $9\mu A/\mathrm{cm}^{2}$ so that the individual 
neurons are identical and oscillatory. 

The synaptic input current from $j$-th neuron is scaled by the synaptic 
strength $\kappa_{ij}$, which changes due to plasticity. 
The adaptation of $\kappa_{ij}$ occurs discontinuously whenever one of the 
neurons $i$ or $j$ spikes. 
More specifically, the discontinuous change is given by the following 
plasticity function
\begin{equation}
\kappa_{ij} \rightarrow 
\begin{cases}
0, & \text{ if } \kappa_{ij}+\delta W(\Delta t_{ij}) < 0\\
\kappa_{ij}+\delta W(\Delta t_{ij}), & \text{ if } 0 \le \kappa_{ij}+\delta 
W(\Delta t_{ij}) \le \kappa_{\max}\\
\kappa_{\max}, & \text{ if } \kappa_{ij}+\delta W(\Delta t_{ij}) > \kappa_{\max}
\end{cases}\label{plasticity}
\end{equation}
where
$\Delta t_{ij}=t_i-t_j$ is the spike time difference between the postsynaptic and 
presynaptic neurons;
$\delta>0$ is a small parameter determining the size of the single update; 
$\kappa_{\max}>0$ is the maximal coupling; and the plasticity function 
\cite{Gerstner1996,Markram1997a,Clopath2010,Bi1998} is 
\begin{equation}
W(\Delta t_{ij})=c_p e^{-\frac{|\Delta t_{ij}|}{\tau_p}}-c_d e^{-\frac{|\Delta 
t_{ij}|}{\tau_d}}\label{plastfunc}
\end{equation}
with positive parameters $c_p, \tau_p, c_d,$ and $\tau_d$. We also assume no 
autapses and set $\kappa_{ii}=0$.

\begin{figure}[h!]
 \centering
  \includegraphics[width={0.5\columnwidth}]{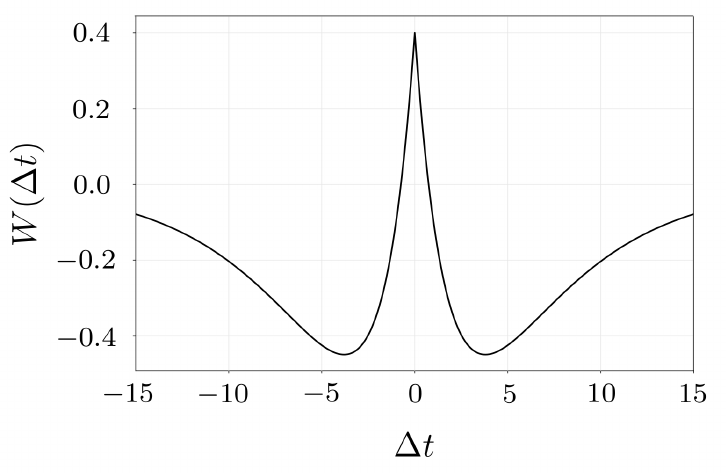}     
  \caption{\textbf{Plasticity function}\\ $W(\Delta t_{ij})$ for $\tau_p=2, \tau_d=5, 
c_p=2, c_d=1.6$\label{fig:plastfunc}}
 \end{figure}

Example of the considered plasticity function $W$ used in our simulations is 
shown in  Fig.~\ref{fig:plastfunc}. This is a symmetric function, which 
corresponds to a potentiation of the coupling weights of the neurons with 
highly correlated firing.  As we will discuss at the end of the results section, there 
is a family of plasticity functions of similar form that allow for the 
frequency clustering.

\section*{Results}

\subsection*{Numerical observation of synchrony and frequency clustering 
\label{sec-numerics}}

\begin{figure}[h!]
 \centering
 \includegraphics[width={\columnwidth}]{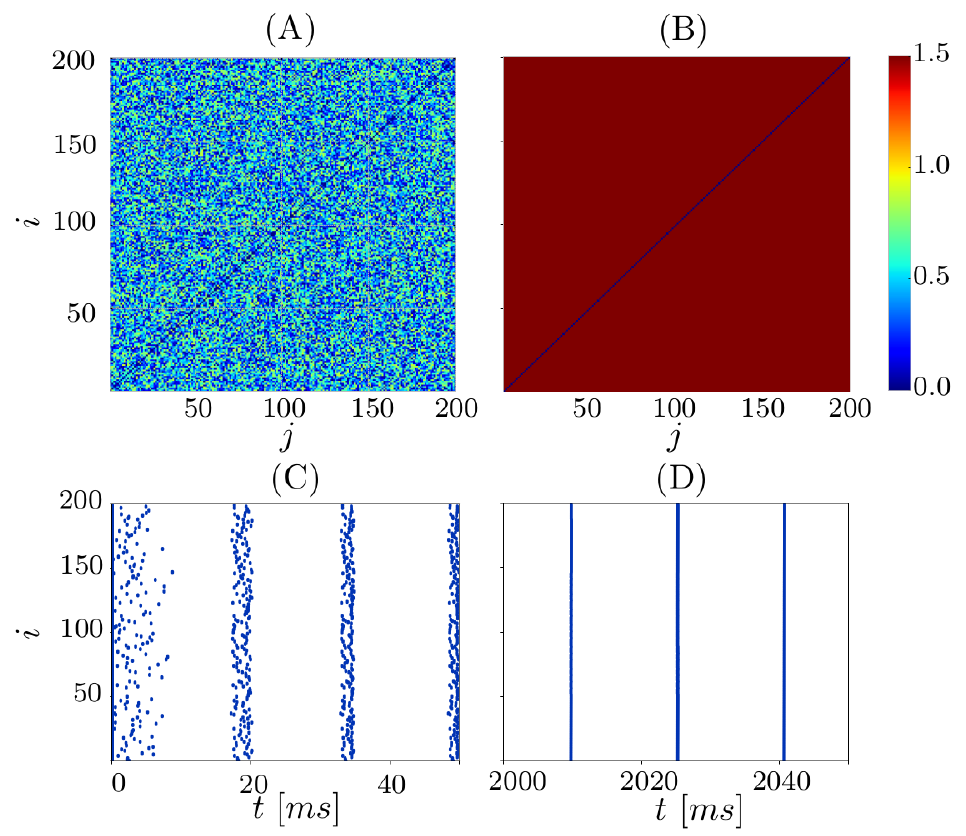}   
  \caption{\textbf{Synchronization into one Cluster}\\
  Evolution of the coupling matrix $\kappa_{ij}(t)$ starting from 
random initial conditions and converging to a completely synchronous state. 
Panel (A) shows initial coupling matrix, (B) the coupling matrix after the 
transient $t=2000$ms. 
  Raster plot of spiking times at the beginning of simulations (C) and after 
the transient (D). The asymptotic state (B,D) is a completely synchronized 
spiking with  all coupling weights $\kappa_{ij}$ potentiated to $k_{\max}$. 
Other parameters   $N=200$, $\tau_p=2$, $\tau_d=5$, $c_p=2$, $c_d=1.6$, and 
$\kappa_{\max}=1.5$.\label{initfusion}}
 \end{figure}

\begin{figure}[h!]
 \centering
  \includegraphics[width={\columnwidth}]{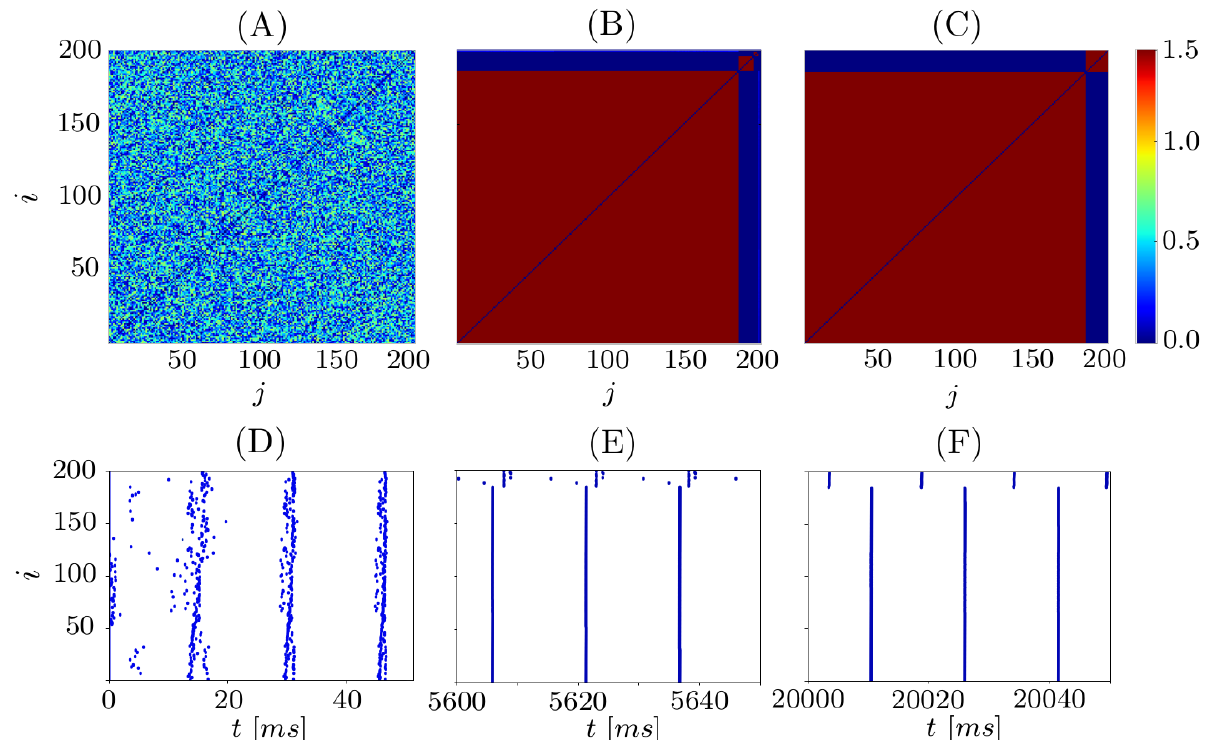}    
  \caption{\textbf{Frequency clusters}\\
  Evolution of the coupling matrix $\kappa_{ij}(t)$ starting from random 
initial conditions and converging to frequency clusters hierarchical in size. Panel (A) 
shows initial coupling matrix, (B) the coupling matrix after the transient 
$t=5600$ms, and (C) $t=20000$ms. 
  (B-F) Corresponding raster plots of spike times. The asymptotic state (C,F) 
is a hierarchical cluster state with  the coupling weights $\kappa_{ij}$ 
potentiated to $k_{\max}$ within each cluster and small or zero otherwise. 
Other parameters  as in Fig.~\ref{initfusion}. The oscillators are ordered accordingly to their mean frequency.
\label{init2cl}}
 \end{figure}

In order to investigate the dynamics of network (\ref{firsteq}), we 
initialize the neurons and the coupling randomly and integrate the system 
numerically. 
For the parameter values $\tau_p=2, \tau_d=5, c_p=2, c_d=1.6$, and 
$\kappa_{\max}=1.5$  we observe two phenomena: complete synchronization and the 
emergence of frequency clusters hierarchical in size, see Figs.~\ref{initfusion} and 
\ref{init2cl}, respectively.

Figure~\ref{initfusion}(A) shows the initial coupling weights, and 
Fig.~\ref{initfusion}(C) illustrates the spike times of the neurons 
at the beginning of the simulation. One can observe that while the neurons 
start with an incoherent spiking, they enhance the coherence already after 
a few spikes due to the interaction between them as well as the plasticity. The plasticity potentiates the connections of neurons that fire together. A complete synchronization is established and the coupling weights increase to $\kappa_{\max}$ after a transient 
(Fig.~\ref{initfusion}(B),(D)). In a completely synchronized 
state, the individual neurons spike simultaneously, 
hence, the spike time differences $\Delta t_{ij}=0$.

The emergence of frequency clusters is shown in Fig.~\ref{init2cl} for two 
clusters. The system in Fig.~\ref{init2cl} possesses the same parameters as in 
Fig.~\ref{initfusion}, and the difference is just another realization of random 
initial conditions. In contrast to the synchronized state, the final 
state shown in Fig.~\ref{init2cl}(F) consist of two groups of synchronized neurons. 
These cluster states also manifest themselves as two groups of strongly coupled 
elements in the coupling matrix $\kappa$ (Fig.~\ref{init2cl}(C)). 
The coupling weights between the neurons from the different groups 
is very small or zero. 

\begin{figure}[h!]
 \centering
\includegraphics[width={0.6\columnwidth}]{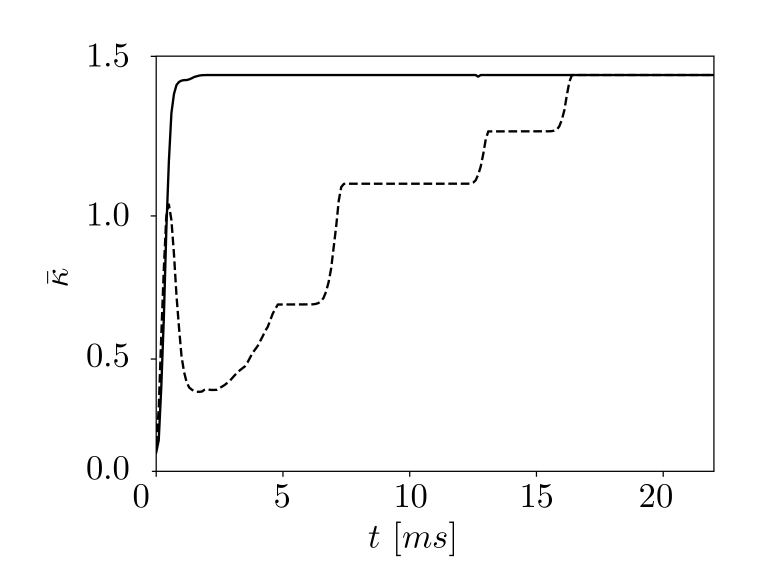}    
  \caption{\textbf{Cluster formation}\\
  Formation of individual clusters over time (corresponds to the dynamical scenario in Fig.~\ref{init2cl}). The dashed 
  and solid curves depict the time course of the mean coupling within the small  
  and big clusters, respectively. }
  \label{coupling}
 \end{figure}

We observe that the largest cluster is formed rather quickly as time evolves, 
whereas the formation of the small cluster takes much more time. This is  
illustrated in Fig.~\ref{coupling}, where the time courses of the mean
coupling within each of the two clusters are shown. 
The average coupling within the big cluster reaches its maximum fast 
(at $t\approx 1000$, solid curve in Fig.~\ref{coupling}), whereas the smaller cluster 
in Fig.~\ref{init2cl}(C, F) is formed through the merging of transient 
clusters and finally establishes at $t\approx 17000$ 
(dashed curve in Fig.~\ref{coupling}). 

\begin{figure}[h!]
 \centering
  \includegraphics[width={0.75\columnwidth}]{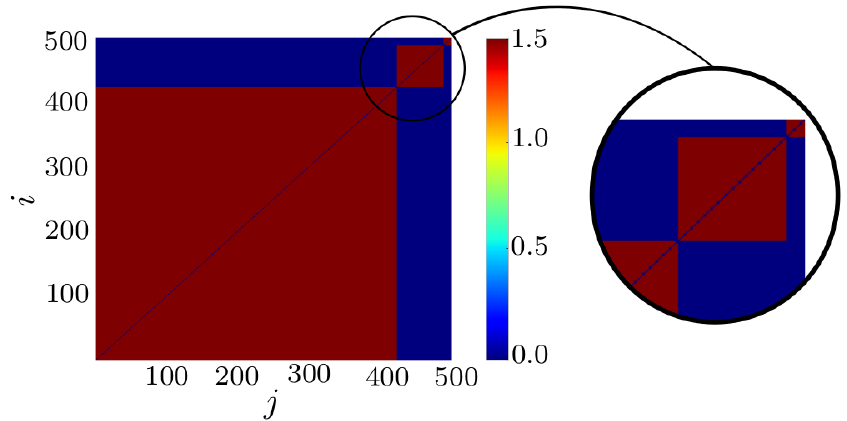}    
  \caption{\textbf{Three-cluster state} Example of a three-cluster state for $N=500$, $\tau_p=2$, $\tau_d=5$, $c_p=2$, 
$c_d=1.6$, and $\kappa_{\max}=1.5$ with a random initial distribution of 
$\kappa_{ij}$ in $[0,0.75]$.
\label{3clusters}}
 \end{figure}

For the states with more clusters, each new formed cluster is significantly 
smaller than the previous one, see Fig.~\ref{3clusters}, where three clusters are shown. 
The spiking period of the cluster appears to be proportional to its size: the 
bigger the cluster the larger is the period. 
Simulation of the cases with even more clusters becomes 
computationally expensive due to large transients.

\subsubsection*{Clustering with independent random input}
\begin{figure}[h!]
 \centering
  \includegraphics[width={\textwidth}]{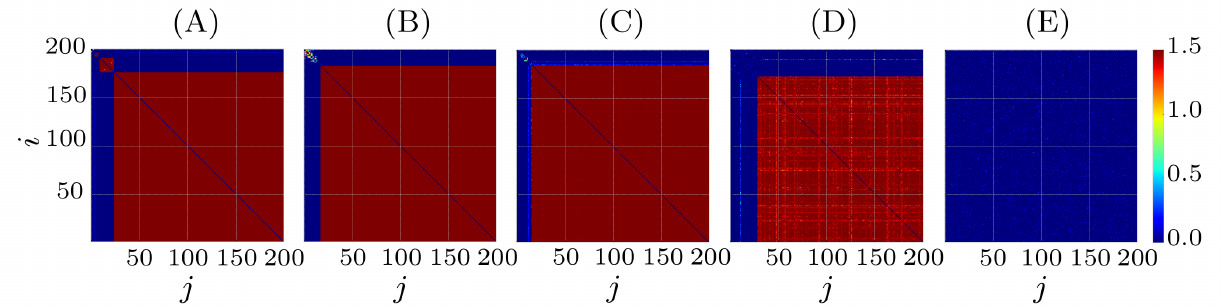}    
  \caption{\textbf{Influence of independent random input on clusters}\\
  Coupling matrices for $t=10000$ms and different amplitudes of independent random input  $I$ (see Eq.~(\ref{alpha})). (A) $I=0.005$, (B) $I=0.01$, (C) $I=0.02$, (D) $I=0.05$ and (E) I=$0.07$. All other parameters as in Fig. \ref{3clusters}
.\label{noise}}
 \end{figure}

To investigate the robustness of our findings, we added an $\alpha$-train as additional independent random input to the membrane potential $V_i$ of every neuron:
\begin{equation}
    I_i^{input}(t)=I(V_r-V_i(t))\sum_{\tau_{i,k}<t}\alpha(t-\tau_{i,k})e^{-\alpha(t-\tau_{i,k})})\label{alpha}
\end{equation}
The Eq.~(\ref{alpha}) models a postsynaptic potential (PSP) that is received by the neuron at certain random times $\tau_{i,k}$. The inter-spike interval is Gaussian distributed $\tau_{i,k+1}-\tau_{i,k}=\Delta\tau_{i,k}\sim \mathcal{N}(14ms,4ms)$. $\alpha$ is set to ${24}/{\langle \Delta \tau_{i,k} \rangle}. $\\
The numerical simulations Fig.~\ref{noise} show that the clustering is still observed under the influence of random input $I_i^{input}(t)$ of intensity $I$. More specifically, for sufficiently weak perturbations with $I<0.01$, all three clusters survive (Fig. \ref{noise}(A)). With increasing the amplitude  $I$,  the clusters start to decouple. The smaller clusters are affected first (Fig. \ref{noise}(B-D)), they start desynchronizing at $I=0.01$. The biggest cluster keeps shrinking in size while $I$ is increased and finally for $I=0.07$  the whole network decouples (Fig. \ref{noise}(E)).

\subsection*{Two clusters in more detail \label{sec-twoclusters}}

In this section we numerically show that depending on the relative 
size of the two clusters, such two-cluster states can be either dynamically stable 
or transient leading to complete synchronization. 
In order to investigate the cluster stability, we initialize the system in a 
two cluster state with the number of neurons $N_s$ in the small  cluster and 
$N_b=N-N_s$ in the big cluster. The total number of neurons is set to $N=50$. 
The inter-cluster couplings are set to zero initially while the intra-cluster 
couplings equals $\kappa_{\max}$. All neurons in the same cluster are 
initialized with the same initial conditions, so the clusters are fully 
synchronized at $t=0$.

\begin{figure}[h!]
\centering
\includegraphics[width={\textwidth}]{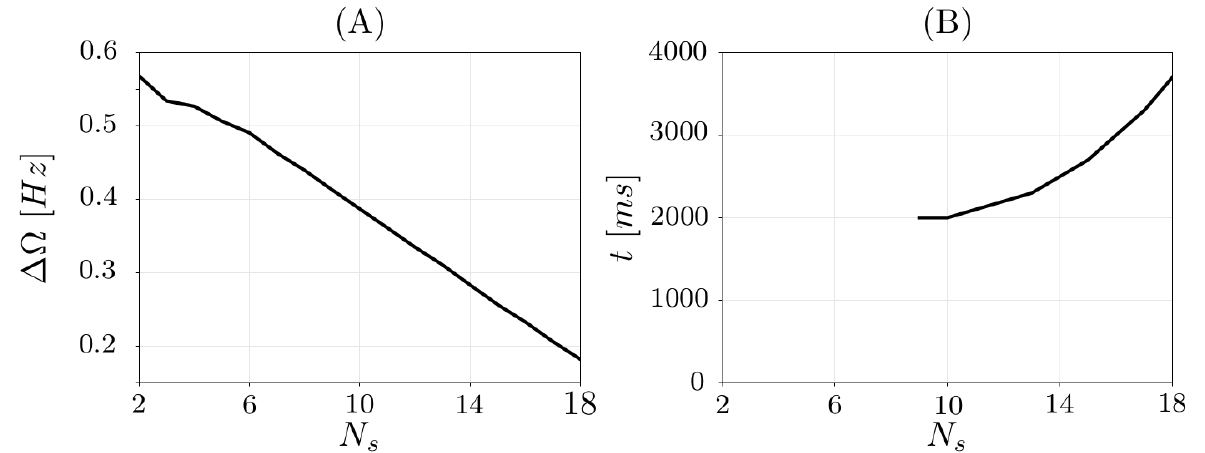}
\caption{\textbf{Cluster frequencies and time until fusion}\\ \label{Change} (A) Difference between synchronization frequencies 
of the two clusters for different size of the smaller cluster $N_s$. 
(B) Time until cluster fusion for different initial size of the smaller 
cluster $N_s$.}
\end{figure}

\begin{figure}[h!]
\centering{ \includegraphics[width={\columnwidth}]{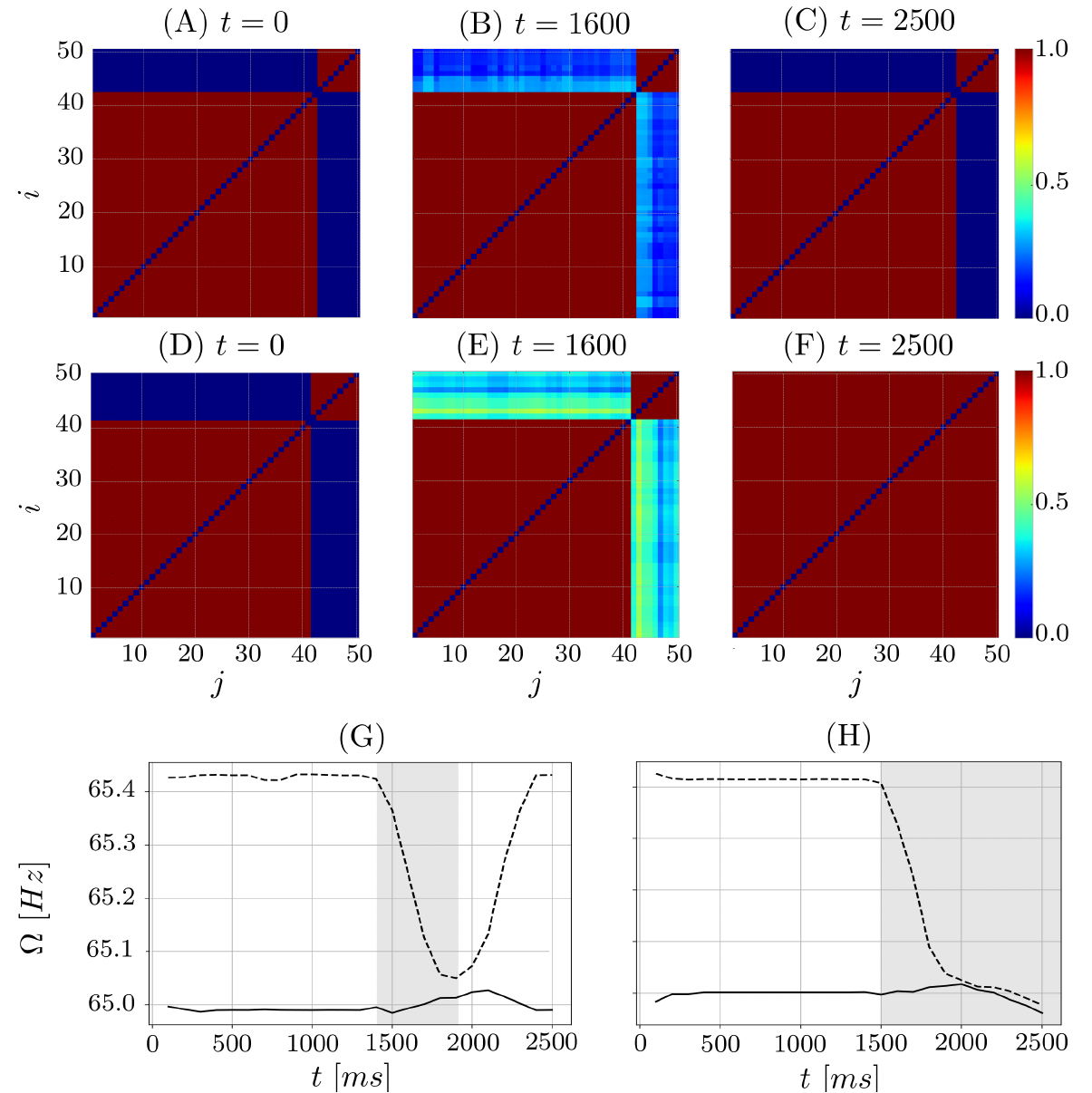}}  
\caption{\textbf{Two cases: Fusion and stable clusters}\\Evolution of the coupling matrix for $N=50$ and the number of neurons 
$N_s=8$ (A)-(C) and $N_s=9$ (D)-(F) in the small cluster. In panels (A)-(C) the 
clusters are stable, while in (D)-(F) they are merging to one synchronous 
cluster. (G, H) Time courses of the spiking synchronization 
frequencies of small ($N_s$ neurons) and large ($N_b$ neurons) clusters 
depicted by dashed and solid curves, respectively, for 
(G) $N_s=8$ and  $N_b=42$ and  (H) $N_s=9$ and $N_b=41$. 
Parameter $\kappa_{max}=1.0$.} \label{Trans1}
\end{figure}

Figure~\ref{Change}(A) shows frequency difference of two uncoupled clusters as 
a function of the size of the small cluster. The frequency difference demonstrates an 
almost linear dependence on the cluster size and decays as the size of the smaller 
cluster increases. Moreover, we also observe that clusters with sufficiently 
different sizes are stable while the clusters of similar sizes, 
in the considered case with $N_s > 8$, are transient, merge into a single 
cluster and eventually lead to a stable completely synchronous state, 
see Fig.~\ref{Trans1}. 

Although the threshold of how different the clusters should be in order to be stable is certainly model dependent, the synchronization of similar clusters is a general property.
The merging of two clusters can be  explained qualitatively as follows. 
Initially uncoupled clusters evolve each with their natural frequencies 
$\Omega_s$ and $\Omega_b$. If their sizes are different, then $ \Omega_s\ne 
\Omega_b$, and the clusters arrive in-phase periodically with the  'beating' 
frequency $\Delta \Omega = \Omega_b - \Omega_s$. As soon as such an in-phase 
episode occurs, the interspike intervals $\Delta t_{ij}$ between any two 
neurons from different clusters become small and, hence, due to the plasticity 
rule  $W$  (see Eq.~\eqref{plasticity} and Fig.~\ref{fig:plastfunc}) the 
inter-cluster coupling weights increase. Moreover, the duration of such an 
in-phase episode depends on the frequency difference between the clusters. As a 
result, for a small frequency difference $\Delta \Omega$, the time interval 
where the clusters are practically in-phase is sufficiently long in order to 
potentiate the coupling weights to their maximum value. This unites the 
two clusters into one. In contrast, for large $\Delta \Omega$, such an episode 
is short, and the inter-cluster coupling remains small, which keeps 
the clusters oscillating at different frequencies in a stable manner.   

Figure~\ref{Trans1}(A)-(C) displays an example of two-cluster stable state with 
$N_s=8$. Starting from  the two-cluster state, after $t=1500$ms,  the coupling 
between the clusters increases, see Fig.~\ref{Trans1}(B)  due to the 
''in-phase episode'' when the clusters are synchronous. Afterwards, 
however, the inter-cluster coupling weights return to their initial 
configuration (Fig.~\ref{Trans1}(C)), since the spike time differences 
for neurons from different clusters are again far enough apart to cause 
the depression of the inter-cluster synapses. Such a process is repeated 
every time the clusters meet and is typical for the stable cluster states.
A typical case of transient clusters is presented in Fig.~\ref{Trans1}(D)-(F) 
for $N_s = 9$. The inter-cluster coupling is again potentiated when 
the clusters meet, but it does not decrease again, and the clusters 
merge in a single cluster of a fully coupled and synchronized regime 
(Fig.~\ref{Trans1}(F)). The transient time that could be elapsed 
until the cluster fusion depends on the cluster size 
as illustrated in Fig.~\ref{Change}(B).

Figures~\ref{Trans1}(G, H) show how the spiking 
frequency of the clusters change over time. During the in-phase episode, 
the cluster with the higher natural spiking rate slows down significantly, 
while the slower cluster (with larger number of neurons $N_b$) 
speeds up a little. For a stable cluster state the cluster frequencies 
again deviate from each other (Fig.~\ref{Trans1}(G)), whereas 
all neurons fire with the same frequency when the clusters 
unite into one (Fig.~\ref{Trans1}(H)). We found this phenomenon 
for different numbers of neurons and different $\kappa_{\max}$. 
Increasing $\kappa_{\max}$ increases the initial period 
difference, but the behavior in general stays the same.

\begin{figure}[h!]
 \centering
  \includegraphics[width={\columnwidth}]{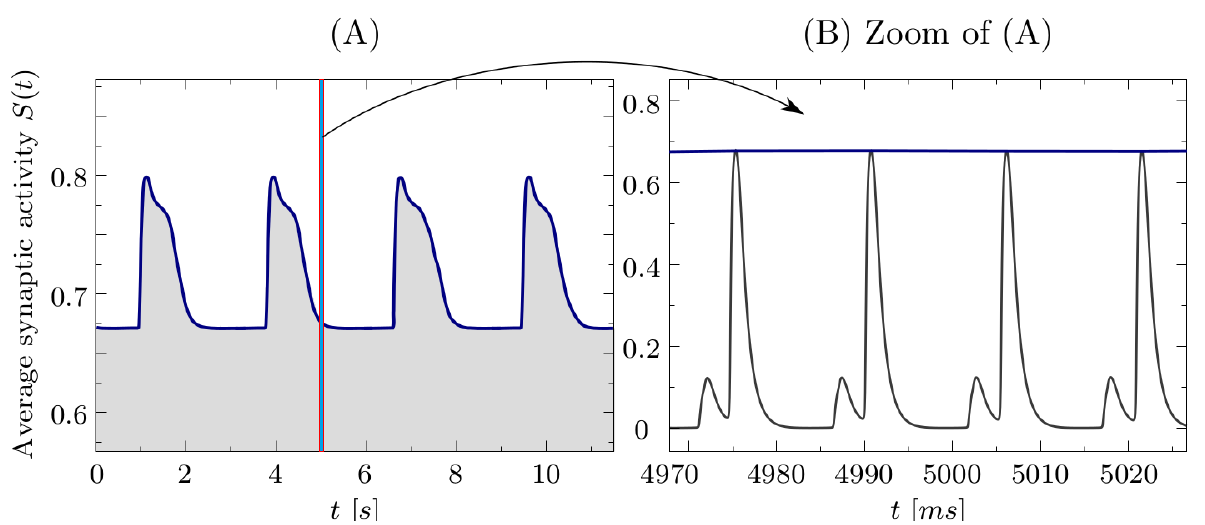}   
  \caption{\textbf{Mean synaptic activity}\\Mean synaptic activity $S(t)$ of the neural population 
  in the case of stable two cluster state. Panel (A) shows the dynamics of $S(t)$ 
  on the time interval of 12\,s, where modulation of the amplitude (blue line) is visible, 
  while the fast oscillations  are not recognized on this timescale. The maximum amplitude corresponds to the two clusters being synchronised, while the low amplitude corresponds to the clusters being out of phase.  Panel (B) shows 
  the zoom of a small time interval. The modulation takes place on the timescale which 
  is two orders of magnitude larger than the individual spikes of $S(t)$ as well 
  as individual neural spikes in both clusters. Cluster frequencies 
  $\omega_1=0.065012$~kHz and $\omega_2=0.065416$~kHz. The corresponding period 
  of modulation is $T\approx 2.5 s$. 
  \label{fig:meansyn}}
 \end{figure}

Figure \ref{fig:meansyn} shows the dynamics of the mean synaptic activity 
$S(t)=\frac{1}{N}\sum_{i=1}^N s_i(t)$ of the network in the case of two stable clusters, 
which models the dynamics of LFP. During the in-phase episodes of the two clusters, 
$S(t)$  has a higher amplitude, because both clusters spike synchronously. 
The maximum amplitude is generated by maximum synchronization in the network.
The low amplitude of $S(t)$, on the other hand, corresponds to the time 
intervals when the clusters are out of phase.  In the latter case, the mean 
synaptic activity shows two peaks, the higher peak is generated by the larger 
cluster and the lower by the smaller one, see Fig.~\ref{fig:meansyn}(B). 
For the considered case, the synchronized oscillations of individual neurons 
in the clusters take place at a time scale of several milliseconds 
(period $\sim 15$~ms, Fig.~\ref{fig:meansyn}(B)), see also Fig.~\ref{Trans1}(G), (H). The neurons are tonically spiking. The frequency difference $\Delta \Omega$ 
between clusters is, however, of the order of sub-Hz, because 
the corresponding cluster frequencies are close to each other (Fig.~\ref{Trans1}(G), (H)).  
Then the modulation of $S(t)$ is observed at a much slower timescale 
of a few seconds, which is of two orders of magnitude slower than 
the intrinsic neural firing, see Fig.~\ref{fig:meansyn}(A), as observed 
in empirical data of the brain activity \cite{Mantini2007,Magri2012}. 

In the following section, a phenomenological model is introduced in order to 
further investigate the dynamics of two clusters. 

\subsection*{Phenomenological Model \label{sec-phenmodel}}

\subsubsection*{Model derivation}

In this section we introduce a reduced qualitative model for the coupling and 
phase difference of two clusters.  
The model is based on the assumption that oscillators are synchronized 
identically within each cluster and the coupling between the clusters is weak. 
As a result, the interaction between oscillatory clusters can be described in 
the framework of two coupled phase oscillators that are interacting via their 
phase differences \cite{Hoppensteadt1997,Pikovsky2001,Guckenheimer1975, 
Winfree2001}
\begin{eqnarray}
\dot \varphi_1 & = & \omega_1 -   F_1(\varphi_1-\varphi_2), \\
\dot \varphi_2 & = & \omega_2 -   F_2(\varphi_2-\varphi_1),
\end{eqnarray}
where $\omega_1$ and $\omega_2$ are the natural frequencies of the individual 
clusters,  $F_1$ and $F_2$  are effective interaction functions. For the 
phase difference $\varphi=\varphi_1-\varphi_2$,  this system reads  
\begin{equation}
\dot{\varphi}=\omega- F(\varphi) \label{phi}
\end{equation}
where  $\omega=\omega_1-\omega_2$  is the difference of the natural frequencies, 
and $F(\varphi) = F_1(\varphi) - F_2(-\varphi) $.  

Since the clusters are synchronized for a sufficiently small frequency mismatch $\omega$, the  periodic interaction function $F(\varphi)$ must satisfy $F(0)=0$  and  $F'(0)>0$. The latter means that there is a stable equilibrium 
$\varphi=0$ for small $\omega$. 
Aiming at a qualitative insight, we further simplify the model by assuming that 
$F(\varphi)=\sigma \sin(\varphi+\alpha)$, where $\sin\varphi$ can be viewed as 
a first Fourier harmonic of the interaction function and $\sigma$ as an 
effective coupling weight. The parameter 
$\alpha=\sin^{-1}(\omega/\sigma_{\max})$ is a constant phase shift assuring that 
the phase difference of the synchronized cluster is zero. In fact, for small $\omega$, this parameter is also small and it does not play important role in the qualitative behavior of the model apart from a small shift of the synchronized state to $\varphi=0$.

\begin{figure}[h!]
 \centering
  \includegraphics[width={\textwidth}]{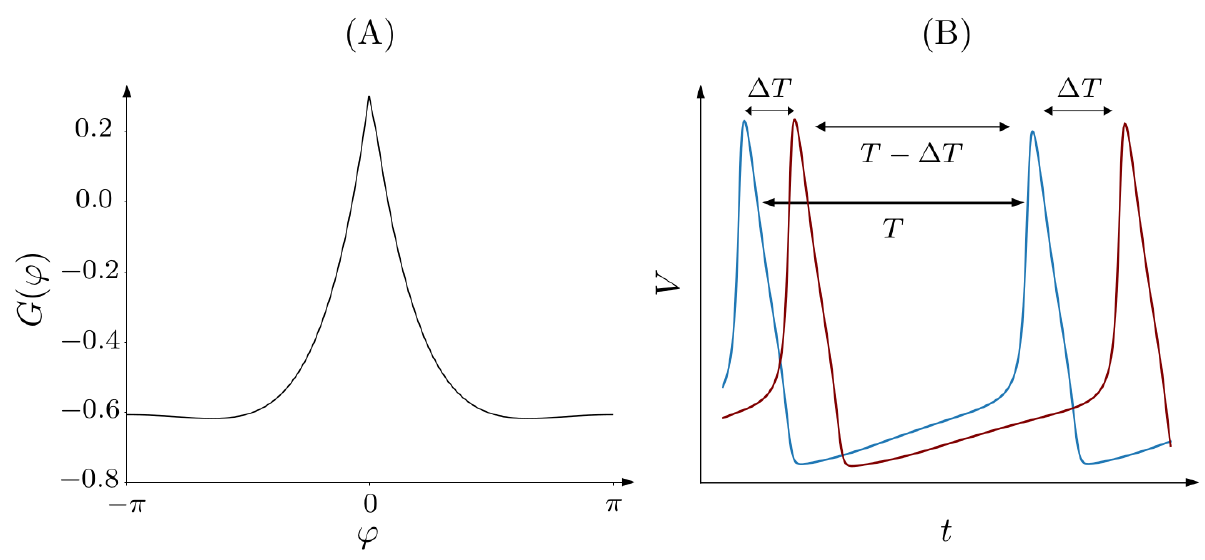}     
  \caption{\textbf{Update function G}\\(A) Update function $G(\varphi)$ for $\tau_p=2, \tau_d=5, c_p=2$, and
$c_d=1.6$. 
  (B) Schematic spiking of two oscillators with spike time difference $\Delta T$ 
and periods close to $T$.
  \label{G}}
 \end{figure}

Another component of the model is the plasticity-driven changes of the coupling 
$\sigma$. 
In order to derive the equation for $\sigma$, we consider the STDP update in 
the case of a periodic motion of the clusters. 
We assume that the coupling $\sigma$ is proportional to an averaged coupling 
between the clusters. This is a natural assumption in the case of weakly 
coupled systems. Let us find out how the update of the intercluster coupling 
depends on the phase difference $\varphi$. 
For a given phase difference $\varphi$ and the frequencies $\omega_1=\bar 
\omega+\omega/2$, $\omega_2=\bar \omega - \omega/2$ (here we introduced the 
mean frequency $\bar \omega$), 
the spiking period of the both clusters can be approximated as 
$T \approx 2\pi/\bar \omega $ up to  small terms of order $\omega$, 
and 
the distance $\Delta T$ between the spikes of two clusters 
 $$
 \Delta T = \left[ T \frac{\varphi_1}{2\pi} - T \frac{\varphi_2}{2\pi} \right] 
\; \mathrm{ mod }\; T 
 = \left[ T \frac{\varphi}{2\pi}\right] \; \mathrm{ mod }\; T 
 \approx \frac{\varphi \; \mathrm{ mod }\; 2\pi}{\bar \omega}.
 $$
Since the spike time differences $\Delta T$ and $T-\Delta T$ occur 
recursively, see Fig.~\ref{G}(B), the updates per unit time  sum to the 
function  
\begin{equation}
\label{Geq}
  \frac{\delta}{T} (W(T-\Delta T)+W(\Delta T))=
\frac{\delta\bar\omega}{2\pi} 
 G(\varphi),
\end{equation}
where 
\begin{equation}
    \label{Gfinal}
 G  (\varphi):= W\left(\frac{2\pi- (\varphi \; \mathrm{ mod }\; 2\pi)}{\bar 
\omega}\right)+W\left(\frac{\varphi \; \mathrm{ mod }\; 2\pi}{\bar 
\omega}\right).
\end{equation}

Since the update of $\sigma$ is proportional to the obtained function, and 
taking into account the smallness of $\delta$, this update  can  be written as 
$\dot \sigma =\varepsilon G(\varphi)$, where $\varepsilon$ is a small parameter of 
the coupling adaptation that controls the scale separation between the fast 
dynamics of the clusters and  the slow dynamics of the coupling.

Additionally, the coupling strength  $\sigma(t) $ should be bounded to the 
interval $[0,\sigma_{\max}]$ by imposing cut-off conditions.
More specifically, the derivative $\dot \sigma$ is discontinuous at the 
boundaries $\sigma=0$ and $\sigma=\sigma_{\max}$, i.e. 
$\dot \sigma = \max \{ 0,  \varepsilon G(\varphi) \} $  for  $\sigma=0 $
and $\dot \sigma = \min \{ 0,  \varepsilon G(\varphi) \} $  for  
$\sigma=\sigma_{\max} $.
The considered cut-off corresponds to ''hard'' bound conditions 
\cite{Song2000}. Another possibility would be ''soft'' or ''multiplicative'' 
bounds \cite{Rubin2001}, when the update is proportional to the distance to the 
boundary. We consider here the hard bound, since it corresponds to the hard 
bound of the STDP rule for HH system. 

The final phenomenological model reads as follows
 \begin{align}
  \dot{\varphi}&=\omega-\sigma \sin(\varphi+\alpha), \label{model21} \\
  \dot \sigma & = \varepsilon \cdot 
    \begin{cases}
     G(\varphi) & \text{for } 0< \sigma < \sigma_{\max} ,\\
    \max \{ 0,  G(\varphi) \}   & \text{for } \sigma=0 ,\\
    \min \{ 0,  G(\varphi) \}   & \text{for } \sigma=\sigma_{\max}.\\
    \end{cases}
\label{model22}
\end{align}
with frequency mismatch  $\omega>0$ and 
$\alpha=\sin^{-1}(\omega/\sigma_{\max})$. 

\subsubsection*{Properties of the model}

\begin{figure}
\centering
 \includegraphics[width={\textwidth}]{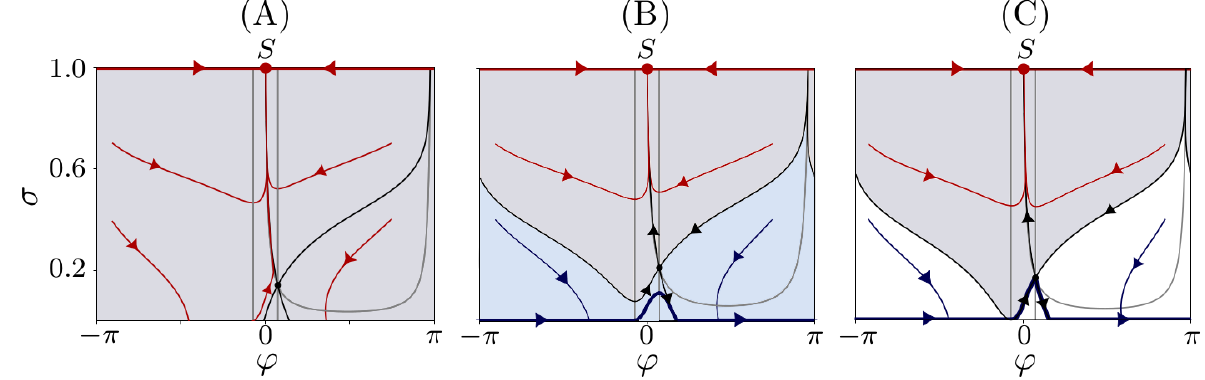}
\caption{\textbf{Phase portrait phenomenological model}\\Phase portraits of model (\ref{model21})-(\ref{model22}) for 
(A) monostable regime of complete synchronization; 
(B) co-existence of stable synchronized and clustered states; 
and (C) bifurcation moment of transition between the phase 
portraits illustrated in (A) and (B). The basins of attraction 
of the synchronized  regime (point $S$), clustered  
state (limit cycle indicated by thick black curve) 
and the saddle fixed point $(\varphi^{\ast}, \sigma^{\ast})$ 
are depicted by gray, 
blue, and white colors, respectively. The nullclines of the 
system and stable and unstable manifolds of the saddle point 
are indicted by the thin gray and black curves, respectively. 
Parameters (A) $\omega=0.037$~kHz, (B) $\omega=0.06$~kHz, 
(C) $\omega\approx 0.455$~Hz, and the other parameters 
$\tau_p=2, \tau_d=5, c_p=2, c_d=1.6$, and $\varepsilon=0.08$.
\label{phaseportraits}}
\end{figure}

Phase space of system (\ref{model21})-(\ref{model22}) is two dimensional with 
$(\varphi,\sigma) \in S^1 \times [ 0,\sigma_{\max} ] $.
The nullclines are given by $G(\varphi)=0$ for $\dot \sigma=0$ and 
$\sigma={\omega}/{\sin(\varphi+\alpha)}$ for $\dot\varphi=0$ in the internal 
points of the phase space.  For the parameter values as in Fig.~\ref{phaseportraits}, the 
$\varphi$-nullcline corresponds to the two lines $\varphi=\varphi^{*}  \approx 
0.23$ and $\varphi = -\varphi^{*}$, while the
$\sigma$-nullcline to a U-shaped nonlinear curve (grey lines in 
Fig.~\ref{phaseportraits}).

There is just one fixed point $(\varphi^{*},\sigma^*=\omega/\sin\varphi^{*})$ 
of saddle-type within the region $\sigma \in (0,\sigma_{\max})$. This point is 
given by the intersection of the nullclines. Figure~\ref{phaseportraits} shows this fixed 
point and its stable and unstable separatrixes (black lines). An additional 
fixed point as well as periodic attractor emerge in system 
(\ref{model21})-(\ref{model22}) due to the non-smoothness at the boundaries. 
More specifically, three situations are observed: 

\textbf{(I)}: One globally stable fixed point $S=(0,\sigma_{\max})$ 
which corresponds to the fusion of the two clusters into one. The coupling 
$\sigma=\sigma_{\max}$ and the phase difference is zero at the fixed point, see 
Fig.~\ref{phaseportraits}.  All orbits are approaching this stable fixed point 
with time. This corresponding phase portrait is shown in 
Fig.~\ref{phaseportraits}(A). 

\textbf{(II)}: Coexistence of the stable fixed point $S=(0,\sigma_{\max})$ and 
a stable periodic orbit, see Fig.~\ref{phaseportraits}(B).
As in the case (I), the fixed point corresponds to the merging of two clusters. 
The periodic orbit corresponds to two simultaneously existing clusters. The 
clusters possess different frequencies and, as a result, the phase difference 
is not bounded and rotate along the circular direction $\varphi$. 
Part of the periodic orbit is located on the boundary $\sigma=0$, i.e. 
vanishing inter-cluster coupling. The coupling $\sigma(t)$ increases between 
$-\varphi^*$ and $\varphi^*$ and decreases otherwise. 
In fact, one can parameterise the coupling $\sigma$ by the phase $\varphi$ on 
the periodic attractor. In the case when $\sigma(\varphi^*)<\sigma^*$, the 
solution returns to the boundary $\sigma=0$, moves along it till the orbit 
reaches the point $(-\varphi^*,0)$, and the periodic motion repeats.

\textbf{(III)}:
When $\max_\varphi G(\varphi) <0 $ then there exists globally stable 
periodic solution $\varphi=\omega t + \varphi_0$, $\sigma=0$. 
In such a case, the fixed 
point on the boundary disappears. Formally, this corresponds to an uncoupling 
between the clusters.  However, in the original HH system, this parameter 
regime corresponds to complete uncoupling of all oscillators because of the 
depression of all synapses. 

In fact, the parameter boundary between the cases (I) and (II) is determined by 
the condition $\sigma(\varphi^*)=\sigma^*$,
which can be interpreted geometrically as hitting the point $(-\varphi^*,0)$ by 
the stable manifold of the saddle equilibrium point, see Fig.~\ref{phaseportraits}(C).
In this special case, the saddle equilibrium attracts the whole set of points from 
the phase space that is below the stable manifolds, see white area in 
Fig.~\ref{phaseportraits}(C). 
In case (II), the separation between the basins of attraction of the fixed 
point and the periodic orbit are given by the saddle equilibrium and its stable 
manifolds. 
A sufficient condition for the case (III) is given by  $c_d \geq c_p$ and $\tau_d \geq \tau_p$. Under these conditions $G(\varphi)\leq 0$ for all $\varphi$.

Summarizing, the case (II) corresponds to the situation when clusters are 
stable and do not merge into one. For this, initial conditions must belong to 
the basin of attraction of the periodic solution (Fig.~\ref{phaseportraits}(B), 
blue domain). The analysis of the phenomenological model indicates that the cluster 
case always coexists with stable complete synchronization. 

\subsubsection*{Comparison of the model and cluster dynamics in HH network}

In order to compare dynamics of the phenomenological model 
(\ref{model21})-(\ref{model22}) and the original system 
(\ref{firsteq})-(\ref{plasticity}), we ran a series of simulations of the HH 
network for parameter values that allow for a stable two-cluster solution. 
The phases of the clusters are calculated as
$\varphi_{1,2}(t)=2\pi\frac{t-t_{k}}{t_{{k+1}}-t_{k}}+2\pi k \text{ for } t 
\in[t_{k}, t_{{k+1}})$, where $\{t_{1},...,t_{n},...\}$ are spiking times with 
$t_{k}<t_{{k+1}}$ \cite{Pikovsky2001}. Correspondingly, the phase difference is 
$\varphi_{HH}(t)=\varphi_1(t)-\varphi_2(t)$. The coupling measure $\sigma_{HH}$ 
 is given by the mean inter-cluster coupling.

\begin{figure}[h!]
 \centering
   \includegraphics[width={0.7\columnwidth}]{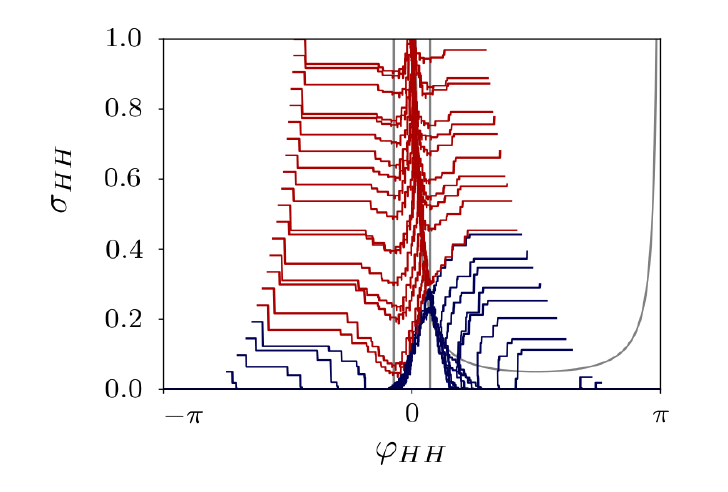}   
  \caption{\textbf{Phase portrait Hodgkin-Huxley model} \\Dynamics of the phase difference between 
  the clusters $\varphi_{HH}$ and mean inter-cluster coupling 
  $\sigma_{HH}$ for the solutions of the HH system 
(\ref{firsteq})-(\ref{plastfunc}) for different initial conditions. 
  $N=50$ with $N_s=7$ neurons in the small cluster and $N_b=43$ in the big one. 
Red orbits converge to the regime of complete synchronization, and blue trajectories 
lead to a stable two-cluster solutions. The nullclines of 
the phenomenological model are shown in gray. 
Other parameters: $\tau_p=2$, $\tau_d=5$, $c_p=2$, $c_d=1.6$, and 
$\kappa_{\max}=1.5$.
  \label{HHPS}}
 \end{figure}

Extracting the quantities $\sigma_{HH}$ and $\varphi_{HH}$ from the numerically 
computed solutions of HH system (\ref{firsteq})-(\ref{plastfunc}) we obtain a 
two-dimensional projection of the solution to the plane 
$(\varphi_{HH},\sigma_{HH})$, see Fig.~\ref{HHPS}. The discontinuities in the 
orbits are related to the discrete STDP updates. Additionally, since the phases $\varphi_{1,2}(t)$ can be firstly  accessed  after the both clusters fired, some of the area of the phase diagram (see white area in Fig.~\ref{HHPS}) was not accessible. This ''empty'' area corresponds to anti-phase initial conditions, which are very sensitive,  and, after each cluster fires, they appear immediately either in the red or blue area.
Nevertheless, the behavior has the same  qualitative features as 
in the phenomenological model, compare Figs.~\ref{phaseportraits} and \ref{HHPS}.

\subsubsection*{Criteria for the emergence of clusters}
 
Model (\ref{model21})-(\ref{model22}) can be used to describe plasticity 
functions, which lead to multiple clusters. 
For this, we investigate numerically the condition 
$\sigma(\varphi^*)=\sigma^*$. More specifically, system 
(\ref{model21})-(\ref{model22}) was initialized at the point $(-\varphi^*,0)$ 
and numerically integrated forward in time. If  $\sigma(\varphi^*)<\sigma^*$, 
the two clusters are stable and do not merge. This procedure can be repeated for 
different parameter values. 

\begin{figure}
\centering
 \includegraphics[width={\textwidth}]{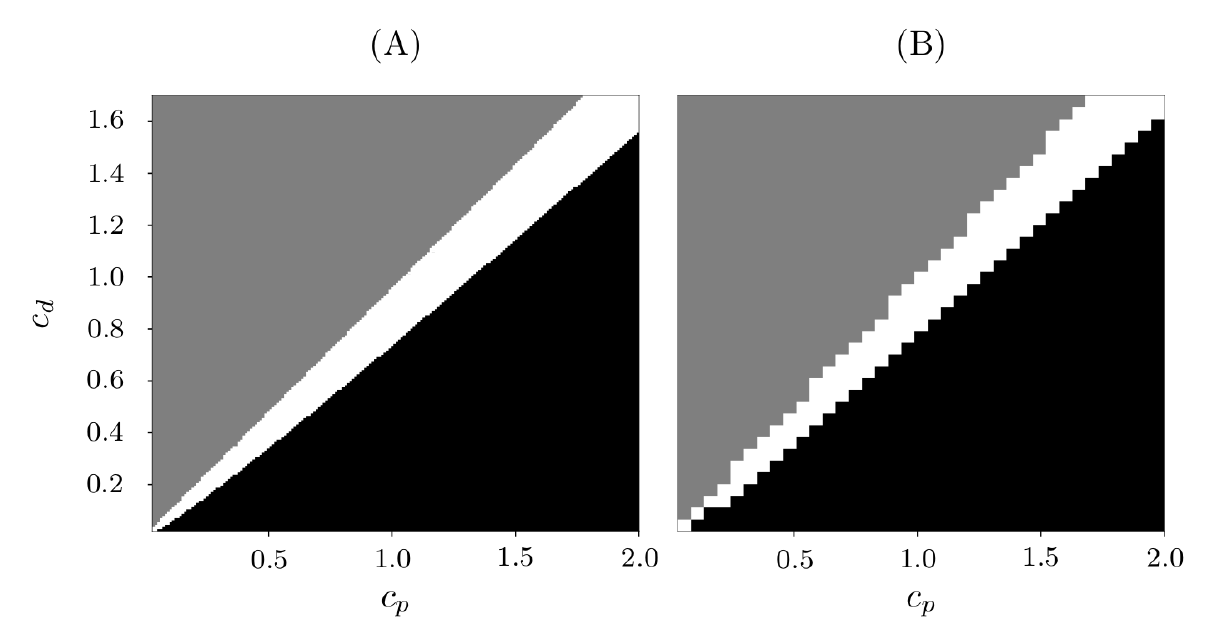}
\caption{\textbf{Parameter $(c_p, c_d)$-plane of the plasticity function}\\
 Panel (A): system 
(\ref{model21})-(\ref{model22}). White region: stable periodic solution 
coexisting with a stable fixed point, case II. Black region: globally stable 
fixed point, case I. Grey region: globally stable periodic solution with 
$\sigma=0$.
Panel (B): original system (\ref{firsteq})-(\ref{plastfunc}). White: stable 
two-clusters (white); black: stable synchrony and no stable clusters; grey: 
decoupling of all neurons. Other parameters $\tau_p=2$, $\tau_d=5$, $N=50$, 
$N_s=7$, and $\kappa_{\max}=1$.
\label{areas}}
\end{figure}

In order to restrict the set of plasticity parameters, we fix $\tau_p=2$ and 
$\tau_d=5$ and vary $c_p$ and $c_d$. The results of the simulation are shown in 
Fig.~\ref{areas}(A). The white, black and grey parameter areas correspond to 
the appearance of stable periodic solution of (\ref{model21})-(\ref{model22}) 
(case (II)), globally stable 
fixed point (case (I)) and the case (III), respectively.

In order to compare the parameter regions obtained for the 
phenomenological model (Fig.~\ref{areas}(A)) with those for the original 
HH system, we ran numerical simulations of system (\ref{firsteq})-(\ref{plastfunc}) 
with $N=50$ neurons and $N_s=7$ neurons in the small cluster. Starting from 
the two-cluster state, we monitor the dynamics of the clusters. Figure~\ref{areas}(B) 
shows the results: the white region corresponds to the case when the clusters survive and 
stay apart after the simulation time $3000$~ms,  black - when the clusters merge 
into one synchronous group, and grey - when the clusters split into uncoupled 
neurons. This behaviour stays qualitatively the same for different cluster 
sizes. However, depending on the frequency difference between the clusters, the 
set of parameters allowing stable cluster states may change its size.

Comparison of the results for the phenomenological system and the HH system in 
the Figs.~\ref{areas}(A,B) shows that the phenomenological model provides a 
reasonable approximation. 

\section*{Conclusion}

Our results show that adaptive neural networks are able to generate 
self-consistently dynamics with different frequency bands. In our case, 
each cluster corresponds to a strongly connected component with a fixed 
frequency. Due to a sufficiently large difference of the cluster sizes and frequencies, 
the inter-cluster interactions are depreciated, while the intra-cluster 
interactions are potentiated. 
In this study, we describe the mechanisms behind the formation and 
stabilization of these clusters. In particular, we explain why the 
significant difference between the cluster sizes is 
important for the decoupling of the clusters. 
From a larger perspective, the decoupling of the clusters in our case is 
analogous to the decoupling of timescales in systems with multiple timescales.

Furthermore, we present a two-dimensional phenomenological model which allows for a 
detailed study of the clustering mechanisms. Despite of the approximations made 
by the derivation, the model coincides surprisingly well with the adaptive Hodgkin- Huxley network. Using the phenomenological model, we find parameter regions 
of the plasticity function, where stable frequency clustering can be 
observed. 

Clustering behavior also emerges at the brain scale, where synchronized 
communities of brain regions constituting large distributed functional networks 
can intermittently be formed and dissolved \cite{Deco2009a,Ponce-Alvarez2015}. 
Such clustering dynamics can shape the structured spontaneous brain activity at rest as measured by fMRI.
In this study, we show that slow oscillations 
based on the modulation of synchronized neural activity can already 
be formed at the resolution level of a single neural population 
if adaptive synapses are taken into account.
These modulations of the amplitude of the mean 
field can be generated in a stable manner [Fig.\ref{fig:meansyn}], 
see also Ref.~\cite{Popovych2015}.
The mechanism relies on fluctuations 
of the extent of synchronization of tonically firing neurons. This is caused
by the splitting of the neural population into clusters and the corresponding 
cluster dynamics.
It might contribute to the emergence of slow brain rhythms 
of electrical (LFP, EEG) and metabolic (BOLD) brain activity reported by 
\cite{Mantini2007,Monto2008,Magri2012,Alvarado2014}. 

However, other mechanisms for generating slow oscillations are possible.
The papers \cite{Bazhenov2002,Compte2003} discussed the
emergence of slow oscillatory activity ($<$ 1Hz) that can be observed
in vivo in the cortex during slow-wave sleep, under anesthesia or
in vitro in neural populations. The suggested mechanism
relies on the corresponding modulation of the firing of individual
neurons, and the slow oscillation at the population level
was proposed to be the result of very slow bursting of individual
neurons that synchronize across the neural population. In contrast, 
the present work shows that the slow oscillations of the population
mean field can also emerge when the firing of individual neurons is not affected. 
The neurons may tonically fire at high frequencies. The amplitude of the  
population mean field then oscillates at much lower frequencies due to the slow 
modulation caused by the cluster dynamics.

Additionally, we would like to mention that the observed frequency clustering resembles phenomenologically the weak chimera states \cite{Ashwin-Burylko,Bick-Ashwin} where clusters with different frequencies are formed in symmetrically coupled oscillators without adaptation. However the properties and mechanisms of the appearance of such clusters are different from those presented here, which are essentially based on the slow adaptation.   

To conclude, we observe self-organised emergence of clusters in neural networks with STDP. The clustering splits the neural population into groups synchronised at different frequencies, which determine the dynamics of the clusters. These cluster dynamics might play a role in low frequency oscillations during the resting state and can be described by a two-dimensional model.

\section*{Acknowledgments}
VR, SY and RB acknowledge the financial support by the Deutsche Forschungsgemeinschaft 
(DFG, German Research Foundation) - Project 384950143, Project 411803875 and Project 308748074.
OVP acknowledges the support by the European Union’s Horizon 2020 Research 
and Innovation Programme under Grant Agreement 826421 (VirtualBrainCloud). ELL acknowledges the financial support  by the S\~ao Paulo Research Foundation (processes FAPESP 2016/23398-8 and 2017/13502-5).

%
%
%

\end{document}